\documentclass[apj]{emulateapj}
\usepackage{lscape}
\usepackage{apjfonts}
\usepackage{graphicx}

\newcommand{\rsun}{{\rm \ R_\odot}}

\newcommand\teff{T$_{\rm eff}$}
\newcommand\logg{log $g$}

\newcommand{\arcsecond}{\mbox{\ensuremath{^{\prime\prime}}}}

\def\e{{\rm E}}

\def\kms{{\rm km}\,{\rm s}^{-1}}

\def\max{{\rm max}}

\def\e{{\rm E}}

\begin{document}
\title{A High-Resolution Spectrum of the Extremely Metal-Rich
 Bulge G-Dwarf OGLE-2006-BLG-265\altaffilmark{1}}

\author{
Jennifer~A.~Johnson\altaffilmark{2},
Avishay~Gal-Yam\altaffilmark{3,4},
Douglas~C.~Leonard\altaffilmark{5},
Joshua~D.~Simon\altaffilmark{3},
Andrzej~Udalski\altaffilmark{6,7}
and
Andrew~Gould\altaffilmark{2,8}
}
\altaffiltext{1}
{The data presented herein were obtained at the W. M. Keck Observatory, which
is operated as a scientific partnership among the California Institute
of Technology, the University of California and the National Aeronautics
and Space Administration. The Observatory was made possible by the
generous financial support of the W. M. Keck Foundation.}
\altaffiltext{2}
{Department of Astronomy, Ohio State University,
140 W.\ 18th Ave., Columbus, OH 43210, USA; 
jaj,gould@astronomy.ohio-state.edu}
\altaffiltext{3}
{Department of Astronomy, California Institute of Technology, 
MS 105-24, Pasadena, CA 91125, USA: avishay,jsimon@astro.caltech.edu}
\altaffiltext{4}
{Hubble Fellow}
\altaffiltext{5}
{Department of Astronomy, San Diego State University, San Diego, 
California, 92182, USA; leonard@sciences.sdsu.edu}
\altaffiltext{6}
{Warsaw University Observatory, Al.~Ujazdowskie~4, 00-478~Warszawa, Poland; 
udalski@astrouw.edu.pl}
\altaffiltext{7}
{Optical Gravitational Lens Experiment (OGLE)}
\altaffiltext{8}
{Microlensing Follow Up Network ($\mu$FUN)}
\begin{abstract}

We present an $R=45,000$ Keck spectrum of the microlensed 
Galactic bulge G-dwarf OGLE-2006-BLG-265, which has high ($\sim 60$)
signal-to-noise ratio despite its short (15 min) exposure
time because the source was magnified by $A\sim 135$. 
While it is very metal-rich ([Fe/H]=0.56), 
the higher temperature of this star compared with the luminous
red giants usually measured in the bulge gives its spectrum many
unblended atomic lines. We measure the abundances of 17 elements, 
including the first abundances for S and Cu
in a bulge star. The [$\alpha$/Fe] ratios are subsolar,
while the odd-Z elements are slightly supersolar, trends that are also 
seen in the more metal-rich stars in the Bulge and the local Galactic disk. 
Because the star is a dwarf, the [O/Fe], [Na/Fe], and [Al/Fe] ratios
cannot be attributed to internal mixing, as is sometimes claimed for
giants. 
Similar high-resolution spectra could be obtained for about a dozen
bulge dwarf stars per year by means of well-designed target-of-opportunity
observations.

\end{abstract}


\section{Introduction
\label{sec:intro}}

While high-resolution spectroscopy is the gold standard of stellar analysis,
the daunting photon requirements of this technique generally limit
its application to the brightest stars within a given stellar population, 
i.e., the giants.  The one major exception is stars from the 
local neighborhood, where the dwarfs can partially compensate for 
their dimness by their greater number and hence (statistically) their
greater proximity.  For distant systems, like globular clusters and
the Galactic bulge, high-resolution spectra of dwarfs would require exposure 
times that are at least 100 to 1000 times longer than those of giants,
which in most cases is prohibitive.

Nevertheless, there are compelling reasons to obtain dwarf spectra
in these systems.  Element abundances in dwarfs and giants are not
necessarily the same even when the progenitors of these stars formed
from gas with identical compositions.  Mixing on the giant branch
can destroy some elements and dredge up others.  
Measuring
the surface-composition differences between dwarfs and giants in the
same population is an important test of theories of stellar evolution.
Moreover, even when the dwarf and giant element abundances are identical,
the {\it derived} values may differ if the atmospheric models used
to estimate the abundances are in error.  Hence, comparing dwarf and
giant spectra is also a powerful way to test atmospheric models.
Driven by these considerations, several groups  
(e.g., \citealt{boesgaard:98}),
have, by prodigious efforts,
obtained high-resolution spectra of subgiant and main-sequence
turnoff stars in globular clusters.

High-magnification gravitational microlensing events provide a unique
opportunity to acquire such spectra. Microlensing events are rare,
affecting only about $\tau\sim 10^{-6}$ Galactic bulge stars at any
given time. The fraction of all events with maximum
magnifications $A_\max\gg 1$ is only $A_\max^{-1}$.  Nevertheless,
by monitoring more than a hundred million bulge stars, the
Optical Gravitational Lens Experiment 
(OGLE-III)\footnote{http://www.astrouw.edu.pl/$\sim$~ogle/ogle3/ews/ews.html}
Early Warning System (EWS, \citealt{ews}) has been finding about
 600 microlensing events in each of the last four years, of which
about a dozen per year have $A_\max > 100$.  The Microlensing 
Observations for
Astrophysics\footnote{http://www.massey.ac.nz/$\sim$iabond/alert/alert.html}
collaboration has recently inaugurated the second phase of their
experiment and expects to soon find a comparable number of high-magnification
events.

High-magnification microlensing events remain within half their 
peak magnification for a time interval \citep{einstein36},
\begin{equation}
t_{1/2} = {\sqrt{12}\,t_\e\over A_\max} 
\sim 1\,{\rm day}\,{t_\e\over 30\,{\rm days}}
\biggl({A_\max\over 100}\biggr)^{-1}
\label{eqn:thalf}
\end{equation}
where $t_\e$ is the Einstein timescale of the event.  
Hence, for
typical Einstein timescales ($\sim 30$ days), high-magnification
events will be observable by any given telescope at magnifications
$A> 50$.  At substantially higher magnifications, the number
of events accessible to an observatory falls as $A^{-2}$, one
factor of $A^{-1}$ because the events are intrinsically rarer and a second
such factor because the chance that the event will peak near the
observatory drops.  Nevertheless, there is at least one
example of an $A_\max=3000$ event, OGLE-2004-BLG-343, 
that peaked over the Very Large
Telescopes in Chile and was basically recognized as such several hours
before peak \citep{ob04343}. Therefore, at least in principle, it would
be possible to obtain of order a dozen spectra of dwarf stars in the
Galactic bulge per year, each magnified by a factor $A\sim 100$,
with a few at much higher magnifications. 

\citet{benetti:95} took the first spectrum of a microlensing
event towards the bulge, albeit at low dispersion, but sufficient to
determine the spectral type. \citet{minniti:98} reported the
possible detection of Li in a bulge dwarf. \citet{cavallo:03}
presented a preliminary abundance analysis for six stars 
with high-resolution spectra including
three dwarfs magnified by factors between 2.3 and 30. 
Here we report on the
abundances of 17 elements in a microlensed 
bulge dwarf star based on high-resolution
spectra, including the first measurements of S and Cu.


{\section{Observations}
\label{sec:data}}

OGLE-2006-BLG-265 was alerted as a probable microlensing event 
toward the Galactic bulge (J2000 RA = 18:07:18.9, Dec = $-$27:47:43;
$l=3.39$, $b=-3.55$)
by OGLE EWS on 2006 May 25.  On June 5, OGLE
issued a further alert that this would be a high-magnification
event, with $A>60$.  Intensive photometric observations were then
carried out by several collaborations, including 
the Microlensing Follow Up Network 
($\mu$FUN)\footnote{http://www.astronomy.ohio-state.edu/$\sim$microfun/},
primarily with the aim of searching for planets 
\citep{mao91,griestsafi,ob05071,ob05169}.
Results of that
search will be presented elsewhere.  The event actually peaked
on June 6 (HJD 2453893.238) at $A_\max\sim 220$.  Two of us 
(A.G.-Y., J.D.S)
happened to be at the Keck Telescope when we received the flurry of 
$\mu$FUN emails describing this event.  We interrupted our
normal program to obtain two exposures of this event, which totaled 
15 minutes.
The observations were carried out approximately 3.6
and 2.7 hours before peak,
when the magnification averaged $A=135$.

Using standard microlensing techniques (e.g., \citealt{ob03262}),
$\mu$FUN determined that the dereddened color and magnitude of the
source were $(V-I)_0=0.63\pm 0.05$, $I_0=18.11\pm 0.10$. OGLE independently
found a similar value. The error
is due to possible differential reddening between the microlensed
source and the red clump, which is assumed to have the same 
$(V-I)_0=-1.00$ as the local Hipparcos clump. If (as seems
likely) the source lies at approximately the Galactocentric distance,
then its absolute magnitude is $M_V=4.3$ and its radius is 1.2$\rsun$. 
That is, it is a solar-type star.

We observed OGLE-2006-BLG-265 
using the High Resolution Echelle Spectrometer (HIRES,
\citealt{vogt:94}) mounted on the Keck-I 10m telescope on 2006 June 6 with
a 600s exposure ending at 14:07:11 UT and a 300s second exposure
ending at 14:49:53 UT. These spectra were obtained under good
conditions (clear sky, $<1$\arcsecond seeing). We used a
$0.81$\arcsecond$\times 3.5$\arcsecond slit, yielding a resolution
of $45,000$. We binned by $\times2$ in the spatial direction to increase
the signal-to-noise ratio
(S/N). The data were reduced using the MAKEE package \citep{barlow:02} 
and additional IRAF\footnote{IRAF is distributed by the National
Optical Astronomy Observatories, which are operated by the Association of
Universities for Research in Astronomy, Inc., under cooperative agreement
with the National Science Foundation.} scripts we have developed. The reduction
included flat-fielding using internal flats, wavelength calibrations
using Th-Ar arcs, and MAKEE extraction using the trace of a bright
calibration star. The two spectra were combined with $scombine$ in 
IRAF. We also observed a rapidly rotating hot star for
telluric line removal and a radial velocity (RV) standard. The two
HIRES chips covering the green and red parts of the
spectrum have higher S/N than the chip with the blue part, 
and we therefore focused our analysis on the data from 
5355\AA{} to 8365\AA. There are small gaps in the spectrum, but
only in the reddest orders. The S/N per resolution element is $\sim$60 over
most of the analyzed spectrum. By cross-correlating (with $fxcor$) the spectrum
of OGLE-2006-BLG-265 with the RV standard HD 222563 
(RV$=5.6\,\kms$, \citealt{udry:99}), we found a heliocentric RV of $-154\,\kms$ 
for the microlensed star. This measurement,
together with the source location projected toward the dense fields of
the bulge, supports its identification as a bulge star.

\begin{figure*}
\includegraphics[angle=270,width=6.5in]{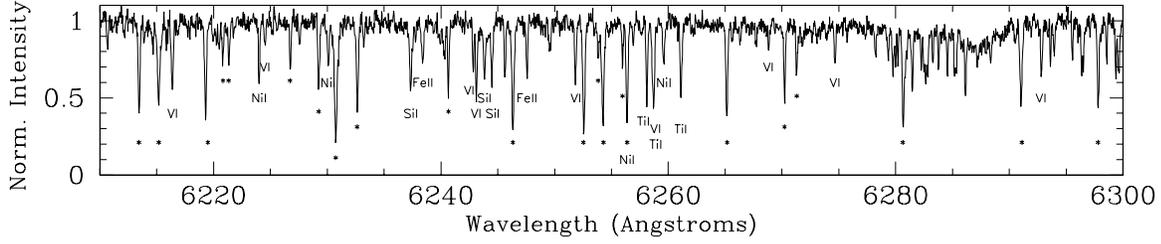}
\caption{A portion of the spectrum of  OGLE-2006-BLG-265. Most of
the strong lines are labeled. The \ion{Fe}{1} lines are marked with
an asterisk (*). The
broad dip at 6280\AA{} is from a diffuse interstellar band, and
there are numerous telluric lines at $\lambda>6275$ \AA.)}

\label{fig:spec}
\end{figure*}

The spectrum clearly shows features consistent with its high
reddening and large distance from the Sun, particularly the
interstellar absorption around the Na D lines and the
presence of the strong diffuse interstellar bands at 5780\AA{}
and 6283\AA{} \citep{herbig:95}. In Figure~\ref{fig:spec} we show part of 
the spectrum.

\vspace{.1in}

\section{Abundance Analysis}
\label{sec:analysis}
We measured equivalent widths with 
SPECTRE\footnote{http://verdi.as.utexas.edu/spectre.html} 
and then translated those into abundances with the 2002
version of MOOG\footnote{http://verdi.as.utexas.edu/moog.html} 
\citep{sneden:73}. We interpolated the Kurucz grid\footnote{http://kurucz.harvard.edu/grids.html} with overshooting for [M/H]$=0.5$ (the limit of
the Kurucz grid) for 
our model atmosphere. We found the 3-$\sigma$ limit for
Li by performing a $\chi^2$ test with the data and synthetic spectra
from MOOG.

\subsection{Model Atmosphere Parameters}
We set \teff{} by requiring that there be no trend in derived
abundance in the \ion{Fe}{1} lines with excitation potential.
We next adopted the
\logg{} for which the average Fe abundance from the \ion{Fe}{1} and
the \ion{Fe}{2} lines was the same to within 0.01 dex. 
We derived the
microturbulent velocity $\xi$ by demanding no correlation between the 
\ion{Fe}{1} abundance derived from a line and its equivalent width.
Our final values are \teff = 5650K$\pm$150K, \logg= $4.4\pm 0.3$,
$\xi=1.2\pm 0.3\,\kms$ and [M/H]$=0.50\pm 0.15$,
where the error in \logg{} is essentially all due to uncertainty in \teff. 
We checked our parameters for consistency with other information
in two ways. First, we combined the spectroscopic \teff{} with the 
microlens-estimated radius to find the luminosity and then converted
this to mass using the zero-age main sequence 
Padua isochrones for Z=0.030 \citep{girardi:02}.  This yielded
\logg{} = 4.34, in good agreement with our
spectroscopic value.   Second, we converted the spectroscopic 
\teff\ to color
\citep{alonso:96}, which yielded $(V-I)_0=0.705\pm0.04$, 
i.e. within $1.2\,\sigma$ of the microlens-estimated value.

\subsection{Atomic Data}
We adopted the line lists from \citet{bensby:03,bensby:04} for 
almost all lines. The exceptions are the \ion{Mn}{1}, 
\ion{Sc}{2}, \ion{V}{1}, \ion{Co}{1} and 
\ion{Ba}{2} and some \ion{Cr}{1} lines, 
for which we used the VALD \citep{vald} values. We used the hyperfine splitting linelists for
Mn, V, Sc, Co, Cu and Ba from \citet{johnson:06} and references therein.
We adopted the solar system isotopic ratios for the Cu and Ba isotopes. 
The enhancement factors to the \citet{unsold:55} damping constants 
from both Bensby et al. papers were used for elements they studied. Otherwise, 
the Uns\"old values were adopted. The NLTE corrections of the \ion{O}{1} triplet
were taken from \citet{bensby:04}. Finally, we used the 
photospheric abundances of \citet{grevsauv} as the solar values.

\subsection{Error Analysis} We calculated errors for both the
$\log\epsilon({\rm X})\equiv\log(N_{\rm X}/N_{\rm H}) + 12.0$
 and [X/Fe] values using the general method described in 
\citet{mcwilliam:95}. Correlations between \teff{} and \logg{}, 
[m/H] and \teff, \logg, and $\xi$ were taken into account, while the correlations
of other combinations of parameters are negligible in our analysis.
We found the random error in our analysis due to uncertainties in 
equivalent widths and atomic data by measuring the standard error
of the sample ($\sigma$) for elements with multiple lines and dividing
by the square root of the number of lines. For elements with
one or two lines, we calculated the error in the equivalent
width from the formula in \citet{johnson:06} and adopted a generous error
in the oscillator strength of 0.1 dex for the random error.

\section{Results}
\begin{figure*}
\includegraphics[width=6.5in]{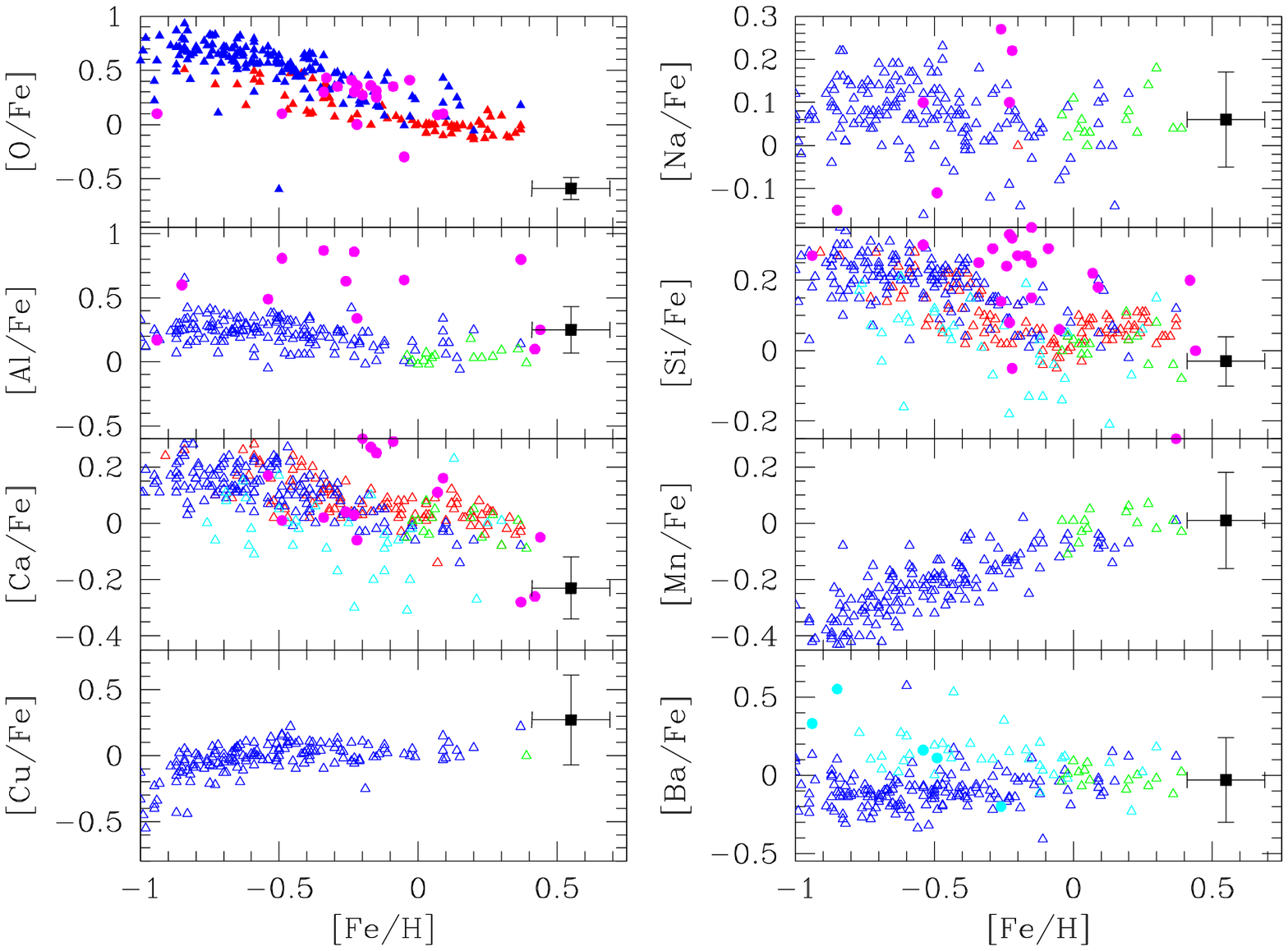}
\caption{Abundance ratios for OGLE-2006-BLG-265 (black) compared with
local samples of thin and thick disk stars from 
\citet{bensby:03,bensby:04,bensby:05} (red), \citet{reddy:03,reddy:06} (blue),
and \citet{chen:03} (green), 
bulge giants from \citet{mcwilliam:94} and \citet{rich:05} (magenta),
and ``bulge-like'' nearby
stars from \citet{pompeia:02,pompeia:03} (cyan).}
\label{fig:abund}
\end{figure*}

The abundances and their errors ($\sigma_{\epsilon}$ and $\sigma_{[X/Fe]}$) 
are summarized in Table~\ref{Tab:Abund}. 
In Figure~\ref{fig:abund},
we plot the abundances compared with recent literature values for
several populations in the Milky Way. 
We note three overall patterns. First, OGLE-2006-BLG-265 is potentially
the most metal-rich star known, although first a careful differential analysis
with other extremely metal-rich stars from the Bulge \citep{fulbright:06} and 
from the local neighborhood (e.g., \citealt{santos:05}) would need
to be done.
Second, element ratios
with O, Si, Ca, Mn, Cu and Ba follow the trends seen in the local
disk/bulge. Third, the light elements Na and Al are similar
in this dwarf to the bulge giants measured by \citet{mcwilliam:94} with
the exception of the large Al-enhancements seen in a few giants. The low
[O/Fe] abundance measured in the lensed dwarf is a reflection of its
initial values of these elements and is 
clearly not the result of internal processing, as
could be the case with giants. We also measured a 3-$\sigma$ upper limit
for Li of log$\epsilon<2.08$ dex.

\begin{deluxetable}{lrrrrrr}
\tablenum{1}\label{Tab:Abund}
\tablewidth{0pt}
\tablecaption{Abundances}
\tablehead{
\colhead{Ion} & \colhead {log$\epsilon$} &  \colhead{$\sigma_{\epsilon}$} 
&\colhead{[X/Fe]} & \colhead{$\sigma_{[X/Fe]}$} & \colhead{$\sigma$}
& \colhead{N$_{\rm lines}$} 
}
\startdata
\ion{O}{1}    & 8.86 &0.10&$-$0.59 &0.18& 0.10 & 3 \\
\ion{Na}{1}   & 6.95 &0.11&$+$0.06 &0.14& 0.19 & 4 \\
\ion{Mg}{1}   & 8.06 &0.19&$-$0.08 &0.19& 0.17 & 1\\
\ion{Al}{1}   & 7.28 &0.16&$+$0.25 &0.16& 0.20 & 2\\
\ion{Si}{1}   & 8.08 &0.07&$-$0.03 &0.10& 0.16 &14\\
\ion{S}{1}    & 7.61 &0.36&$-$0.28 &0.39& 0.36 & 1\\
\ion{Ca}{1}   & 6.69 &0.11&$-$0.23 &0.09& 0.22 &12\\
\ion{Sc}{2}   & 3.90 &0.21&$+$0.17 &0.12& 0.19 & 4\\
\ion{Ti}{1}   & 5.48 &0.19&$-$0.10 &0.10& 0.14 & 6\\
\ion{V}{1}    & 4.65 &0.18&$+$0.09 &0.13& 0.15 & 8\\
\ion{Cr}{1}   & 6.31 &0.13&$+$0.08 &0.07& 0.21 &14\\
\ion{Mn}{1}   & 5.96 &0.17&$+$0.01 &0.10& 0.09 & 6\\
\ion{Fe}{1}   & 8.05 &0.14&$-$0.01 &\nodata& 0.19&98\\
\ion{Fe}{2}   & 8.07 &0.16&$+$0.01 &0.07& 0.16 & 9\\
\ion{Co}{1}   & 5.54 &0.25&$+$0.06 &0.24& 0.22 & 1\\
\ion{Ni}{1}   & 6.87 &0.15&$+$0.06 &0.04& 0.18 &31\\
\ion{Cu}{1}   & 5.04 &0.34&$+$0.27 &0.27& 0.25 & 1\\
\ion{Ba}{2}   & 2.66 &0.27&$-$0.03 &0.15& 0.08 & 3\\
\enddata
\end{deluxetable}


For most elements, the abundance ratios we observe in OGLE-2006-BLG-265 
are expected
for a star of its metallicity if we extrapolate the results of
the canonical chemical evolution models of \citet{timmes:95}. 
In their scenario, the low [O/Fe], [Mg/Fe], [Si/Fe], [Ca/Fe]
and [Ti/Fe] (and possibly [S/Fe]) 
ratios reflect the increasing importance of Fe contributed
by Type Ia supernovae, while the solar or supersolar [Na/Fe], [Al/Fe],
[Mn/Fe] and [Cu/Fe] 
(and possibly [Co/Fe]) are the result of the greater ease of forming
these odd-Z elements in more metal-rich Type II SNe. 
The solar value of [Ba/Fe] is expected as the s-process
contribution from AGB stars keeps up with the Fe production. All of
these conclusions suggest that for this bulge star at [Fe/H]=$+0.56$ dex,
the enrichment history has lasted a considerable period, longer than
for the bulge stars with high [$\alpha$/Fe] values at substantially 
lower metallicities ([Fe/H]$\leq 0.0$ dex) observed by \citet{mcwilliam:94} and
\citet{rich:05}.
The ratio of the iron-peak elements Sc, V, Cr and Ni to Fe
are more unexpected, being elevated in all cases above the solar
value, though by small amounts. Additional stars in the bulge, whether
giants or dwarfs, will be helpful in determining the robustness of
those results. Possible enhancements in the iron-peak elements
have been seen in Galactic populations such as in Sc, V and Co in the thick
disk \citep{prochaska:00} and Cr in the bulge \citep{mcwilliam:94}.
The interpretation of Li abundances is complicated by internal
destruction in stars before joining the main-sequence. Our upper limit
is near the level of Li measured in stars of similar
temperature in the Hyades, a young cluster, and
higher than the upper limits determined for older stars in M67 and
in the Galactic disk \citep{balachandran:95, lambert:04}.

This {\it Letter} shows that unique information can be obtained by taking
advantage of the magnification from microlensing to observe otherwise
unattainable objects. With a dozen spectra a year of bulge dwarf stars,
selected without any abundance bias, we have the opportunity to study the
stellar populations in the bulge in a new way.

\acknowledgments
We acknowledge support from:
Hubble Fellowship HST-HF-01158.01-A [NASA contract NAS-5-2655] (AG-Y);
NSF AST-042758, NASA NNG04GL51G (AG);  NASA HST-AR-10673 [NASA contract NAS5-26555] (DCL); NSF AST-0204908, NASA grant NAG5-12212, Polish MNiSW BST grant (AU); 

We recognize and acknowledge the very significant 
cultural role and reverence that the summit of Mauna Kea has always had within 
the indigenous Hawaiian community.  We are most fortunate to have the 
opportunity to conduct observations from this mountain.

\end{document}